# Top–ology*

Chris Quigg
Fermi National Accelerator Laboratory
Batavia, Illinois 60510 USA

**Abstract**

I compare the anticipated properties of the top quark with what is known experimentally about the heaviest fermion. I review the scientific opportunities that will be presented in the near future by copious samples of top. These include refined tests of electroweak radiative corrections, studies of the nature of top itself, and exploration of electroweak symmetry breaking. I conclude by illustrating top's influence in the everyday world.

## 1 The Top Quark Must Exist

Ever since the existence of the $b$-quark was inferred from the discovery of the $\Upsilon$ family of resonances in 1977 [1], we have been on the lookout for its weak-isospin partner. Once the charge of the $b$-quark was established to be $e_b = -1/3$, it was natural to expect that the missing partner should be the upper member of a doublet, with charge $+2/3$. Completing the weak-isospin doublet by finding the top quark is the most natural way to cancel the anomaly of the $(\nu_\tau, \tau)_L$ doublet and ensure an anomaly-free electroweak theory.

The absence of flavor-changing neutral currents in $b$ decays, which would lead to significant branching fractions for dilepton channels such as $b \rightarrow s\ell^+\ell^-$, added phenomenological support to the idea that $b$ is a member of a left-handed weak doublet.

More recently, the accumulation of results on the neutral-current interactions of the $b$-quark has made it possible to characterize the $Zb\bar{b}$ vertex and measure the weak isospin of the $b$-quark. Consider a generalization of the $SU(2)_L \otimes U(1)_Y$ theory in which the $b$-quark may carry both left-handed and right-handed weak isospin. The chiral neutral-current couplings can be written as

$$\begin{aligned} L_b &= I_{L3} - e_b \sin^2\theta_W \\ R_b &= I_{R3} - e_b \sin^2\theta_W \quad , \end{aligned} \tag{1}$$

which differ from the standard-model chiral couplings by the presence of $I_{R3}$. Characteristics of the reaction $e^+e^- \rightarrow b\bar{b}$ permit us to determine the values of

*Lecture given at LISHEP95, LAFEX International School of High Energy Physics, Workshop on Heavy Flavors *cbt*, Rio de Janeiro, February 1995. I have revised the text to incorporate information from the CDF and DØ discovery papers.

$I_{L3}$ and $I_{R3}$ directly from experiment. The partial width $\Gamma(Z^0 \to b\bar{b})$ measures the combination $L_b^2 + R_b^2$. On the $Z^0$ resonance, the forward-backward asymmetry $A_{\rm FB}(Z^0 \to b\bar{b})$ measures the combination $(L_b^2 - R_b^2)/(L_b^2 + R_b^2)$. Far below the resonance, where the forward-backward asymmetry is dominated by $\gamma$-$Z$ interference, $A_{\rm FB}(e^+e^- \to b\bar{b})$ measures the combination $L_b - R_b$. The unique overlap of the allowed regions is for $I_{L3} = -1/2$, $I_{R3} = 0$, the standard-model solution.

## 2 Anticipating $m_t$

Little can be said on general theoretical grounds about the masses of new flavors, but interesting constraints arise from consistency requirements and from phenomenological relationships. Imposing the requirement that partial-wave unitarity be respected at the tree level in the reactions

$$Q\bar{Q} \to (W^+W^-, Z^0Z^0, HZ^0, HH) \tag{2}$$

leads to a condition on the heavy-quark mass $m_Q$, which determines the scale $(G_F m_Q^2 \sqrt{2})^{1/2}$ of the $HQ\bar{Q}$ couplings [2]. For the $(t,b)_L$ doublet of heavy quarks, the restriction amounts to

$$|m_t - m_b| \lesssim 550 \text{ GeV}/c^2. \tag{3}$$

This general constraint can be sharpened appreciably by considering radiative corrections to electroweak observables.

Top-quark loops contribute to the renormalization of low-energy observables such as [3]

$$\rho \equiv \frac{M_W^2}{M_Z^2(1-\sin^2\theta_W)} \approx 1 + \frac{3G_F m_t^2}{8\pi^2\sqrt{2}} \; . \tag{4}$$

A recent fit to electroweak observables from LEP and SLC, together with information from neutrino scattering yields the expectation [4]

$$m_t = 178^{+11+18}_{-11-19} \text{ GeV}/c^2, \tag{5}$$

assuming the validity of the standard model with three generations of quarks and leptons.

Many other observables, particularly those related to neutral-meson mixing and $CP$ violation, are sensitive to the top-quark mass. One example, for which we may expect significant progress over the next five years, is the parameter $\epsilon'$ that measures direct $CP$ violation in the $K^0$-$\bar{K}^0$ system. Figure 1 shows the region favored by state-of-the-art calculations as a function of $m_t$. We expect the theoretical uncertainty to shrink as lattice-QCD calculations mature. The values measured by NA31 at CERN and by E731 at Fermilab are plotted at arbitrary values of $m_t$ [6]. Experiments now being prepared may reduce the experimental uncertainty on $\epsilon'/\epsilon$ to $\pm 1 \times 10^{-4}$.

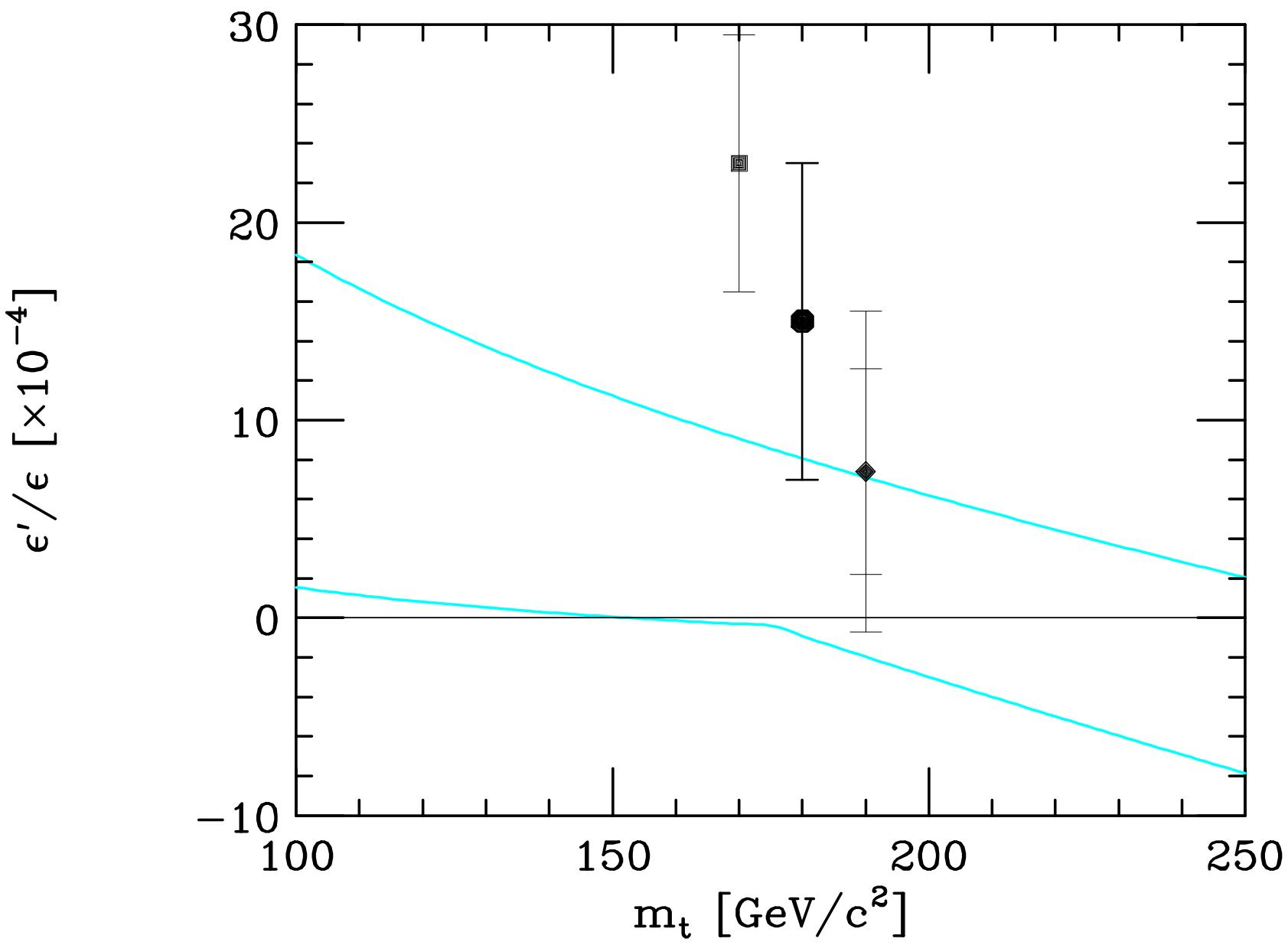

Figure 1: The quantity $\epsilon'/\epsilon$ in the standard model as a function of the top-quark mass. The band shows the region allowed by plausible variations in theoretical parameters. The values measured by NA31 (■) and E731(◆) are shown, together with the world average (●). (G. Buchalla, private communication [5].)

## 3 Production Rates

The calculation of the top-quark production cross section in perturbative QCD has been carried out to next-to-leading order (NLO) and beyond, using resummation techniques [7]. Typical results are shown in Figure 2 for $p^{\pm}p$ collisions at c.m. energies between 1.8 and 14 TeV. At 1.8 TeV, for $m_t = 175$ GeV/$c^2$, the QCD cross section is $\sigma(\bar{p}p \to t\bar{t} +$ anything$) \approx 6$ pb, predominantly from the elementary process $\bar{q}q \to t\bar{t}$. Experts [9] estimate the uncertainty in the calculations as $\pm 30\%$. At 14 TeV, the energy planned for the Large Hadron Collider at CERN, the QCD cross section rises to $\sigma(pp \to t\bar{t} +$ anything$) \approx 800$ pb, predominantly from the mechanism $gg \to t\bar{t}$.

It is interesting to ask what would be gained by raising the top energy of the Tevatron collider by lowering the operating temperature of the superconducting magnets. Figure 3 shows how the top production cross section depends on $m_t$ at $\sqrt{s} = 1.8$ and 2.0 TeV, according to the resummed next-to-leading-order calculation of Laenen, Smith, and van Neerven [7]. For 160 GeV/$c^2 \leq m_t \leq$ 200 GeV/$c^2$, the cross section would increase by about 40% if the c.m. energy were raised to 2 TeV. The fraction of the cross section contributed by $gg$ collisions grows from about 15% to about 20% for a top-quark mass of 175 GeV/$c^2$.

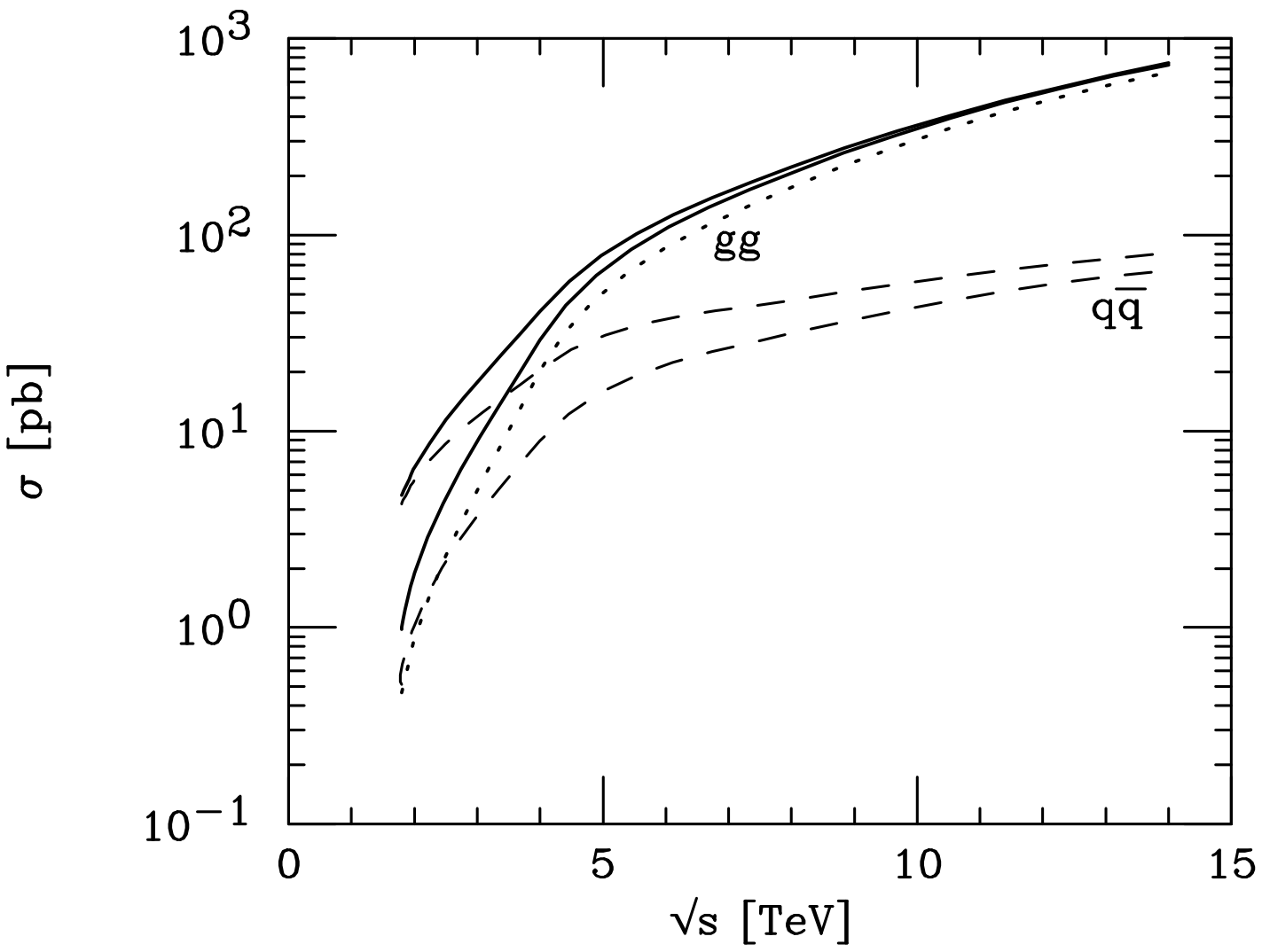

Figure 2: Energy dependence of the cross section for production of 175-GeV/$c^2$ top quarks in $pp$ (lower curves) and $\bar{p}p$ (upper curves) collisions. The contributions of $\bar{q}q$ (dashed curves) and $gg$ (dotted curve) collisions are shown separately. (After Parke, Ref. [8].)

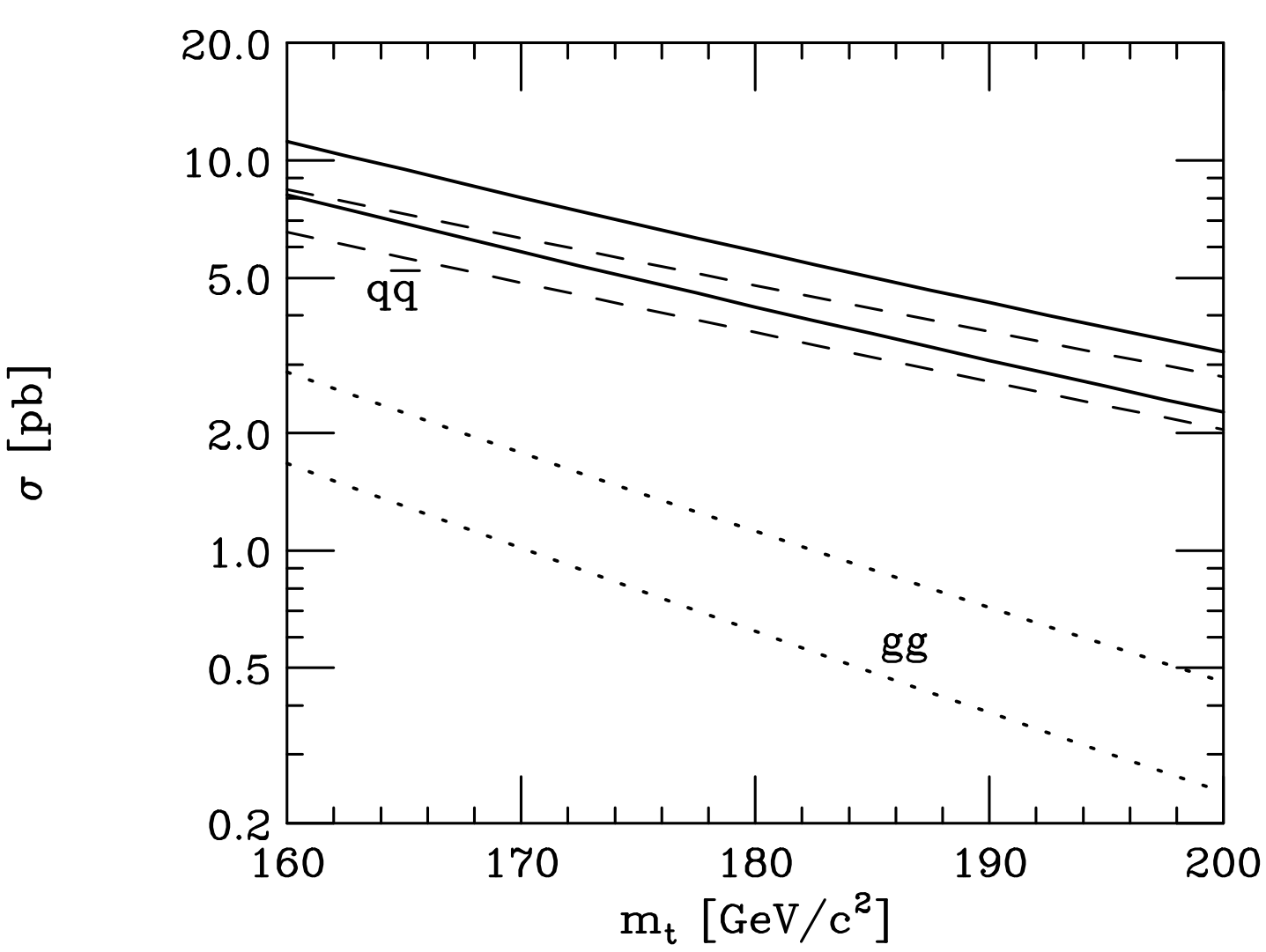

Figure 3: Dependence of the top production cross section, $\sigma(\bar{p}p \rightarrow t\bar{t} +$ anything), upon the top-quark mass in 1.8-TeV (lower curves) and 2.0-TeV (upper curves) $\bar{p}p$ collisions. The standard QCD contributions of $q\bar{q}$ (dashed curves) and $gg$ (dotted curves) are shown separately. (E. Laenen, private communication.)

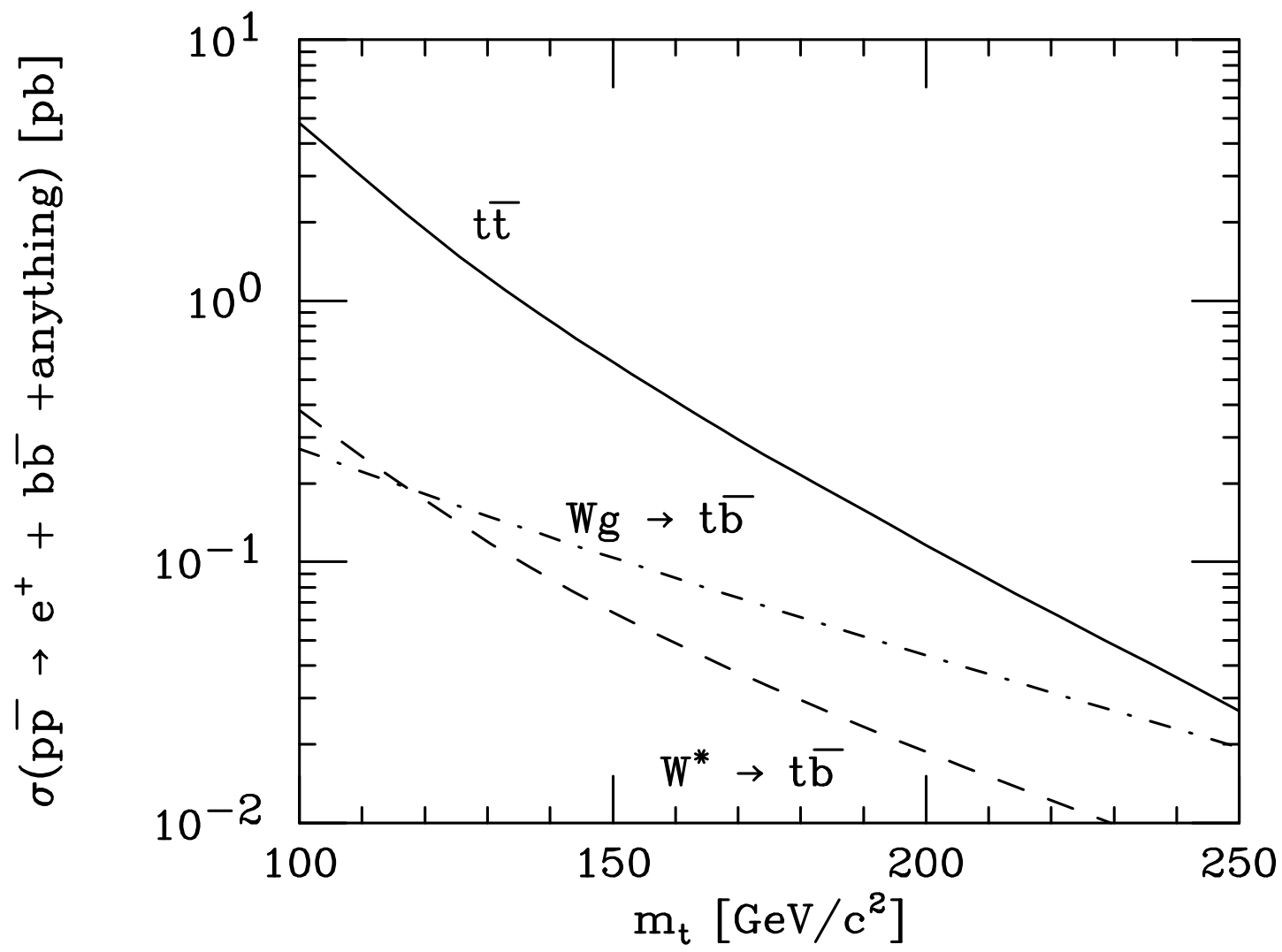

Figure 4: Variation of standard-model contributions to the cross section $\sigma(\bar{p}p \to e^+ b\bar{b} +$ jets) with the top-quark mass at $\sqrt{s} = 1.8$ TeV. Yields are shown for the $t\bar{t}$ (solid curve), $Wg \to t\bar{b}$ (dot-dashed curve), and $W^\star \to t\bar{b}$ (dashed curve) contributions. (After Parke, Ref. [8].)

In addition to the dominant mechanisms for top production included in Figure 3, other conventional sources may take on importance as the integrated luminosity rises. I show in Figure 4 the contributions of the weak-interaction processes $W^+g \to t\bar{b}$ and (virtual) $q\bar{q} \to W^\star \to t\bar{b}$ to the cross section for producing $e^+ +$ jets, assuming that $t \to bW^+$ is top's only decay mode. The final state contains $e^+b\bar{b} + (2,1,0)$ non-$b$-quark jets for the QCD, $W$-gluon, and virtual-$W$ processes. Top has no special standing in the most conventional "new-physics" cases, $W' \to t\bar{b}$ and $Z' \to t\bar{t}$.

## 4 Top Width and Lifetime

In the standard model, the dominant decay of a heavy top quark is the semiweak process $t \to bW^+$, for which the decay rate is [10]

$$\begin{aligned}\Gamma(t \to bW^+) &= \frac{G_F M_W^2}{8\pi\sqrt{2}} \frac{1}{m_t^3} \left[ \frac{(m_t^2 - m_b^2)^2}{M_W^2} + m_t^2 + m_b^2 - 2M_W^2 \right] \\ &\quad \times \sqrt{[m_t^2 - (M_W + m_b)^2][m_t^2 - (M_W - m_b)^2]}. \end{aligned} \tag{6}$$

If there are only three generations of quarks, so that the Cabibbo-Kobayashi-Maskawa matrix element $V_{tb}$ has a magnitude close to unity, then for a top-quark mass of 175 GeV/$c^2$, the partial width is

$$\Gamma(t \to bW^+) \approx 1.55 \text{ GeV}, \tag{7}$$

which corresponds to a top lifetime $\tau_t \approx 0.4 \times 10^{-24}$ s, or 0.4 yoctosecond (ys) [11]. A heavy top is then so ephemeral that there will be no toponium spectroscopy, and indeed no dressed hadronic states containing top. Accordingly, the characteristics of top production and the hadronic environment near top in phase space should be calculable in perturbative QCD [12].

The $t \to bW^+$ decay rate and (partial) lifetime are shown in Figure 5 for a range of top-quark masses.

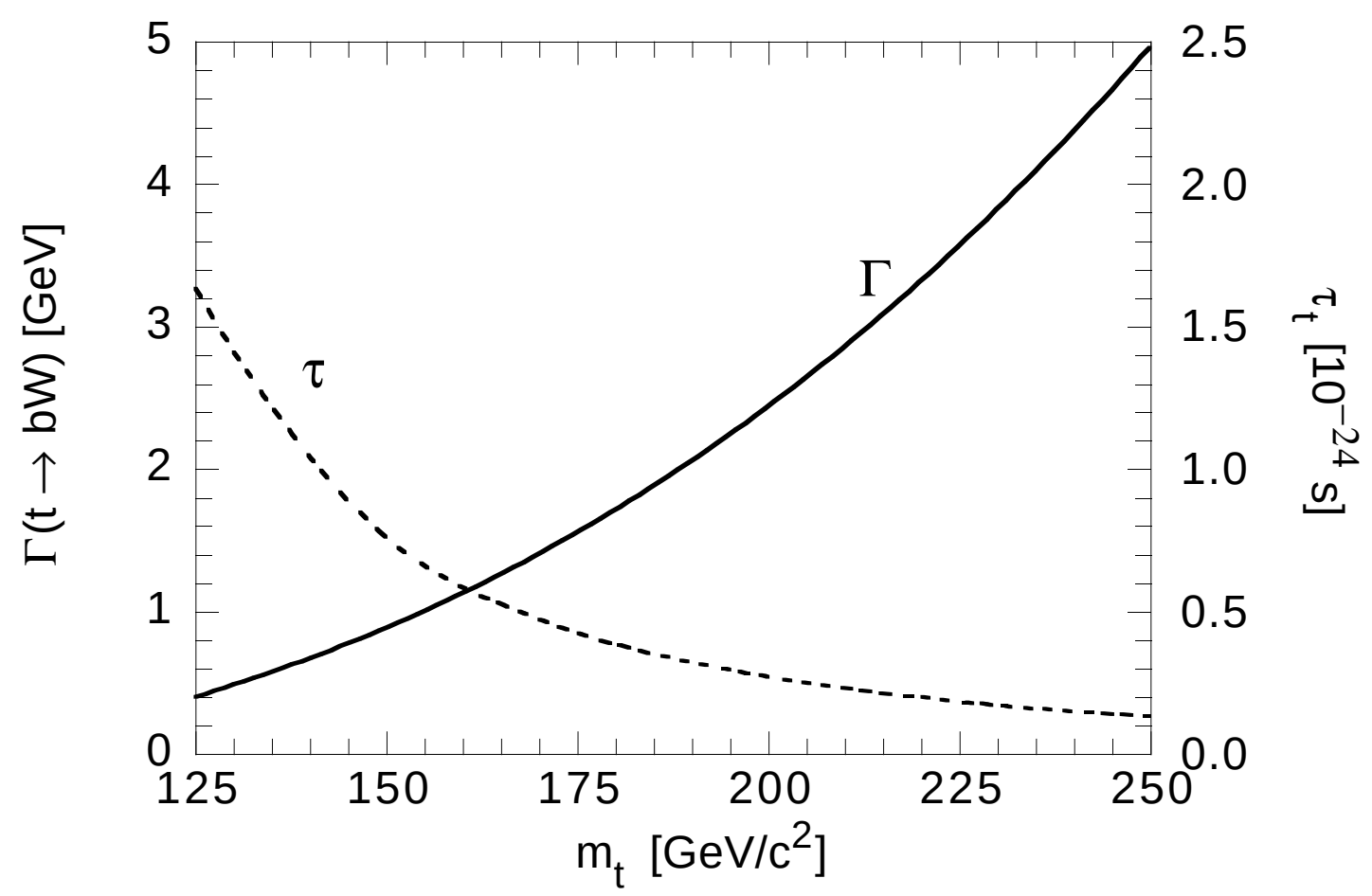

Figure 5: Partial width (solid curve, left-hand scale) for the decay $t \to bW^+$ as a function of $m_t$. The (partial) lifetime is shown as the dashed curve (right-hand scale). A full-strength $t \to b$ transition is assumed.

It is noteworthy that top decay is an excellent source of longitudinally polarized $W$-bosons, which may be particularly sensitive to new physics. Helicity$=-1$ $W$-bosons are emitted with relative weight 1 and helicity$=0$ $W$-bosons with relative weight $m_t^2/2M_W^2$. For $m_t = 180$ GeV/$c^2$, a fraction $f_0 = 71\%$ of the $W$-bosons emitted in top decay will be longitudinally polarized. The decay angular distribution of charged leptons in the $W$ rest frame is

$$\frac{d\Gamma(W^+ \to \ell^+ \nu_\ell)}{d(\cos\theta)} = \frac{3}{8}(1-f_0)(1-\cos\theta)^2 + \frac{3}{4} f_0 \sin^2\theta. \qquad (8)$$

## 5 Discovery of the Top Quark

Shortly after LISHEP95, the CDF [13] and DØ [14] collaborations reported the discovery of the top quark in dilepton and lepton-plus-jets events produced in 1.8-TeV $\bar{p}p$ collisions at the Tevatron. CDF reconstructs a top mass of $m_t = (176 \pm 8 \pm 10)$ GeV/$c^2$ and infers a production cross section $\sigma(\bar{p}p \to t\bar{t} +$ anything$) = 6.8^{+3.6}_{-2.4}$ pb. The DØ result is $m_t = (199^{+19}_{-21} \pm 22)$ GeV/$c^2$, with $\sigma(\bar{p}p \to t\bar{t} +$ anything$) = 6.4 \pm 2.2$ pb.

The observation of top completes the last normal (light-neutrino) generation and provides a crucial parameter of the electroweak theory. The large mass of the top quark suggests that top might stand apart from the other quarks and leptons. Top provides a new window on novel physics through nonstandard production and decay. We now take up some of the implications of the discovery.

# 6 Top in Electroweak Radiative Corrections

The influence of the top quark on electroweak radiative corrections was the basis for the expectations for $m_t$ from precision measurements of electroweak observables. As the top-quark mass is known more precisely from direct measurements, it will be possible to compare predictions for which $m_t$ is an input with other observations. Over the next few years, we can anticipate incisive tests of the electroweak theory from the comparison of the $W$-boson mass with theoretical calculations.

Predictions for $M_W$ as a function of the top-quark mass are shown in Figure 6 for several values of the Higgs-boson mass [15]. I have plotted the CDF measurements of $M_W$ and $m_t$ as representative of the current experimental uncertainties [13, 16]. If $\delta M_W$ improves statistically, it should decrease to about $\pm 80$ MeV by the end of the collider run now in progress. An uncertainty of $\pm 50$ MeV/$c^2$ seems a realistic possibility at both the Tevatron and at LEP200. Improving $\delta m_t$ to $\pm 5$ GeV/$c^2$ in running with the Main Injector at Fermilab will then make for an extremely incisive test of the standard-model prediction. The derived value of $M_W$ to be obtained in the high-statistics neutrino experiment at Fermilab should approach the present precision of the direct $M_W$ measurement. For summaries of what is already known from LEP, see Ref. [3] and [4].

# 7 Is It Standard Top?

The top-quark discovery channels all arise from the production of top-antitop pairs. We expect that all the significant channels will contain a $b\bar{b}$ pair, from the decay chain

$$\begin{array}{l} t\;\bar{t} \\ |\;\;\hookrightarrow \bar{b}W^- \\ \hookrightarrow bW^+\;, \end{array} \tag{9}$$

leading to $b\bar{b}e^{\pm}\mu^{\mp}\nu\nu$, $b\bar{b}e^+e^-\nu\nu$, $b\bar{b}\mu^+\mu^-\nu\nu$, $b\bar{b}\ell\nu$ jet jet.

We expect that decays other than the observed $t \to bW^+$ mode are strongly suppressed. Unless the quark–mixing-matrix element $|V_{tb}| \ll 1$, which could occur if top had a strong coupling to a fourth-generation $b'$ with $m_{b'} > m_t$, the decays $t \to (s,d)W^+$ should be extremely rare [17]. It is important to test this expectation by looking for the rare decays directly, or by comparing the number of observed $(0, 1,$ and $2)$ $b$-tags in a top-quark sample with expectations derived from the various top-production mechanisms and the efficiency for $b$-tagging [18].

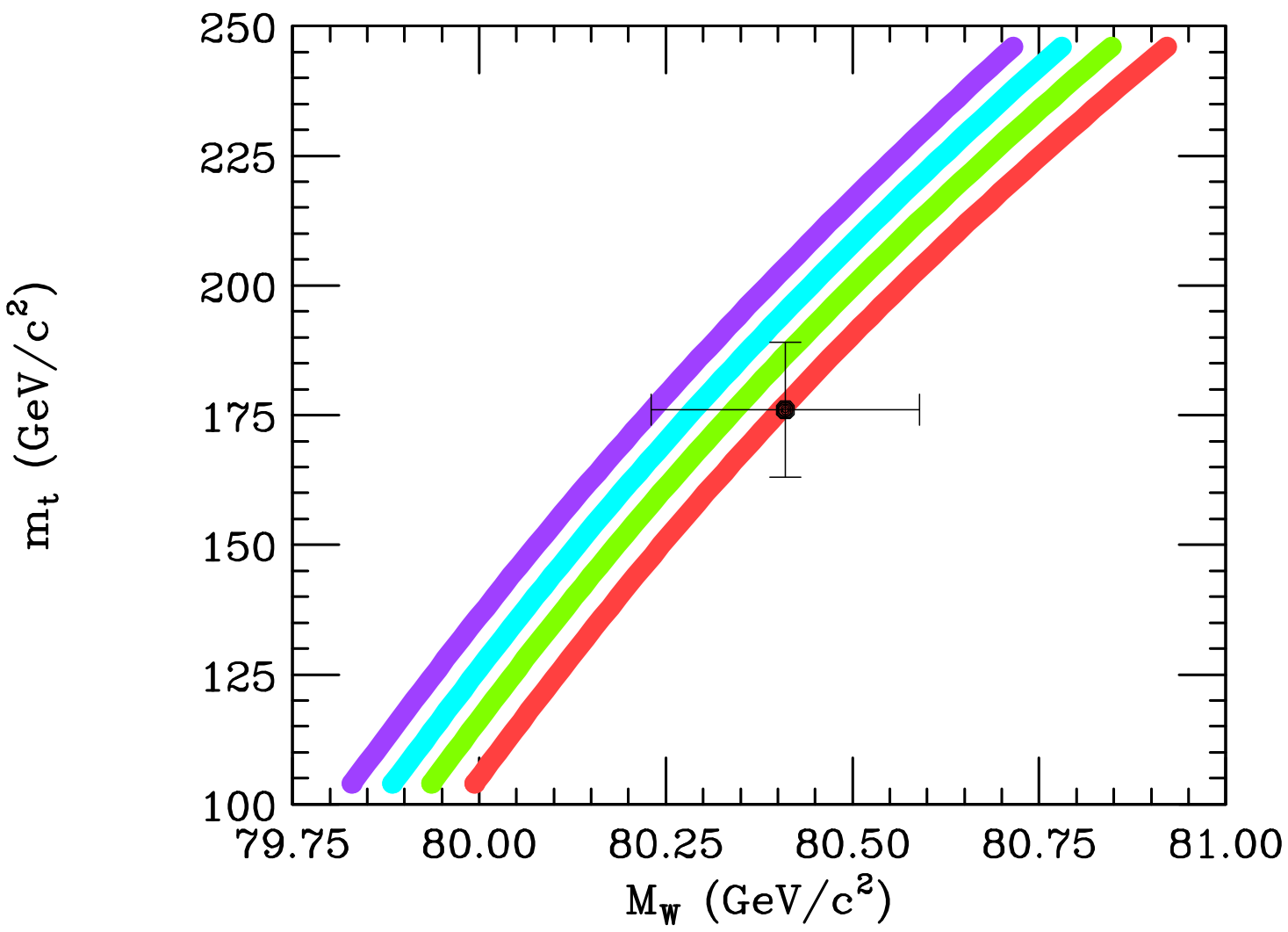

Figure 6: Correlation between the top-quark mass and the $W$-boson mass in the standard electroweak theory [15]. From left to right, the bands correspond to Higgs-boson masses of 1000, 500, 250, and 100 GeV/$c^2$. The thickness of the bands expresses the effect of plausible variations in the value of $\alpha(M_Z^2)$. The point represents the current CDF mass values.

Stelzer and Willenbrock have argued recently that the $W^\star \to t\bar{b}$ process may in time provide the best measurement of the quark mixing-matrix element $|V_{tb}|$ [19]. Prospects for extracting top-quark (and other) parameters from threshold studies at a future $e^+e^-$ linear collider have been surveyed by Fujii, Matsui, and Sumino [20].

The rapid decay of the top quark means that there is no time for the formation of top mesons or baryons. Accordingly, the spin orientation of the top quark at the moment of its production is reflected, without dilution, in the decay angular distribution of its decay products. The lepton angular distribution thus becomes a tool for probing the structure of the charged-current interactions of top [21].

The persistence of the top quark's polarization can be exploited to devise tests for $CP$ violation in top decays. Because the standard model leads to very tiny effects, the top system has great sensitivity to nonstandard sources of $CP$ violation. A brief review of the considerable literature on the subject can be found in Ref. [22].

The branching ratios expected for the flavor-changing neutral-current decays

$$t \to \begin{pmatrix} g \\ Z \\ \gamma \end{pmatrix} + \begin{pmatrix} c \\ u \end{pmatrix} \tag{10}$$

all are unobservably small ($\ll 10^{-10}$) according to the standard electroweak the-ory [23]. Anomalous $Zt\bar{c}$ couplings could lead to a branching fraction as large as a few per cent while respecting current constraints from low-energy phenomenology. High-luminosity experiments at the Tevatron or the LHC should be able to explore branching fractions as small as $\sim 10^{-2}$ and $\sim 10^{-4}$, respectively [24].

Mahlon and Parke have examined the possibility that a threshold enhancement might render observable rare decays like $t \to bWZ$ and $t \to bWH$ [25]. The finite widths of the $W$ and $Z$ bosons do raise the decay rates dramatically near threshold, but the branching fractions remain too small to be observed in the current round of experiments. Over the interval 160 GeV/$c^2 \lesssim m_t \lesssim$ 200 GeV/$c^2$, the branching fraction $\Gamma(t \to bWZ)/\Gamma(t \to bW)$ rises from $1.6 \times 10^{-7}$ to $1.4 \times 10^{-5}$. Detection of this mode in a modest top sample would therefore be a compelling sign of new physics.

Because top is so massive, many decay channels may be open to it, beyond the dominant $t \to bW^+$ mode. The semiweak decay $t \to bP^+$, where $P^+$ is a charged (pseudo)scalar, may occur in multi-Higgs generalizations of the standard model, in supersymmetric models, and in technicolor models. The decay rate $\Gamma(t \to bP^+)$ is generically comparable with $\Gamma(t \to bW^+)$, because both are semiweak [26]. An inferred $t\bar{t}$ production cross section smaller than that predicted by QCD would be a hint that $\Gamma(t \to bW^+)/\Gamma(t \to \text{all}) < 1$, which would argue for the presence of nonstandard decays.

Decay of a charged scalar into fermion pairs would typically proceed at a rate

$$\Gamma(P^+ \to f_i \bar{f}_j) = \frac{G_F p(m_i^2 + m_j^2)}{16\pi} C_{ij}, \tag{11}$$

where $p$ is the momentum of the products in the rest frame of $P^+$ and $C_{ij} = (3, 1)$ for (quarks, leptons). The lifetime of $P^+$ is far too short for it to be observed as a short track: for $M_{P+} = M_W$, $\tau_{P+} \lesssim 10^{-21}$ s (= 1 zeptosecond) [27]. $P^+$ might be reconstructed from its decays into $c\bar{b}$ or $c\bar{s}$, or its presence might be deduced from $P^+ \to \tau^+\nu_\tau$ decays, which would also lead to violations of lepton universality.

The general lesson is that top decays have the potential to surprise. It may therefore be quite rewarding to learn to tag top-bearing events with high efficiency [28].

The large mass of the top quark has an important effect on the pattern of soft-gluon emission from an energetic top. If the energy $\omega$ of the gluon is small compared to the energy $E_Q$ of the heavy quark, $\omega \ll E_Q$, then for a gluon emitted at a small angle $\theta \ll 1$ to the top-quark direction, the angular distribution of the radiation will be

$$d\sigma_{Q \to Qg} \sim \frac{\theta^2 d\theta^2}{(\theta^2 + \theta_0^2)^2} \frac{d\omega}{\omega}, \tag{12}$$

where $\theta_0 = m_Q/E_Q$. For angles larger than the critical value, *i.e.*, for $\theta > \theta_0$, the radiation pattern becomes

$$d\sigma \sim \frac{d\theta^2}{\theta^2} \frac{d\omega}{\omega} \to d(\ln \theta^2) d(\ln \omega), \tag{13}$$

which is doubly logarithmic. Indeed, when $\theta \gg \theta_0$, the emission of successive gluons follows a strict angular ordering, and the multiplicity of hadrons accompanying the heavy quark is the same as it would be for a light quark. In contrast, in the very forward cone defined by $\theta < \theta_0$, there is only a single logarithmic factor, $d\omega/\omega$. In this region, gluon emission is inhibited and the multiplicity of accompanying hadrons is diminished. The regime of reduced multiplicity is known as the dead cone [29].

Although the angular dependence of radiation accompanying a heavy quark has not been measured, there is some evidence for a reduced multiplicity in the number of hadrons emitted by an energetic $b$-quark. The SLD and OPAL experiments have used tagged samples of $e^+e^- \rightarrow Z^0 \rightarrow b\bar{b}$ events to compare the charged multiplicity $\langle n_b \rangle$ of hadrons produced by energetic $b$-quarks with the multiplicity $\langle n_{u,d,s} \rangle$ produced by energetic light quarks [30]. SLD finds $\langle n_b \rangle - \langle n_{u,d,s} \rangle = 3.31 \pm 0.41 \pm 0.79$, while OPAL measures $\langle n_b \rangle - \langle n_{u,d,s} \rangle = 3.02 \pm 0.05 \pm 0.79$. This is a significant suppression of particle emission, and bodes well for the possibility of reconstructing self-tagging $B^{**} \rightarrow B^{(*)}\pi$ decays cleanly for studies of $CP$ violation in $B$ decays. The effect should be considerably larger for top quarks, which will have the added advantage of decaying before they can be dressed into resonances whose effects are not included in the perturbative analysis.

## 8 Top's Yukawa Coupling

In the $SU(2)_L \otimes U(1)_Y$ electroweak theory, Higgs scalars give masses to the electroweak gauge bosons $W^\pm$ and $Z^0$, and also to the elementary fermions. While the gauge-boson masses are determined in terms of the weak mixing parameter $\sin^2\theta_W$, each fermion mass is set by a distinct Yukawa coupling, as

$$m_f = \frac{G_f v}{\sqrt{2}}, \tag{14}$$

where $v = (G_F\sqrt{2})^{-1/2} \approx 246$ GeV. The Yukawa coupling of the electron is $G_e \approx 3 \times 10^{-6}$.

The top quark stands apart from the other fundamental constituents because its Yukawa coupling is very close to unity: $G_t \approx 1$. Is top special? Or is it the only normal fermion?

In either case, the large $Ht\bar{t}$ coupling has several implications. (i) Higgs interactions will exert a significant influence on the evolution of the top-quark mass. As we examine the possibility that the pattern of fermion masses is more intelligible at the unification scale than at the scale of common experience, it is important to evolve the fermion masses to our scale with care [31]. (ii) It is worth reexamining the reactions $q\bar{q} \rightarrow (\gamma^\star, Z^\star) \rightarrow t\bar{t}H$ and $q\bar{q} \rightarrow Z^\star \rightarrow ZH \rightarrow Zt\bar{t}$ at the LHC or, with $e^+e^-$ replacing $q\bar{q}$, at a multi-TeV linear collider. (iii) A heavier top quark is a more important product of heavy-Higgs decay. Figure 7 shows the partial widths for Higgs-boson decay into the dominant $W^+W^-$ and $Z^0Z^0$ channels and into $t\bar{t}$, for $m_t = 175$ GeV/$c^2$. Whether the $t\bar{t}$ mode will be useful to confirm the observation of the Higgs boson, or merely drains probability from the favored $ZZ$ channel, is a question for detailed detector simulations.

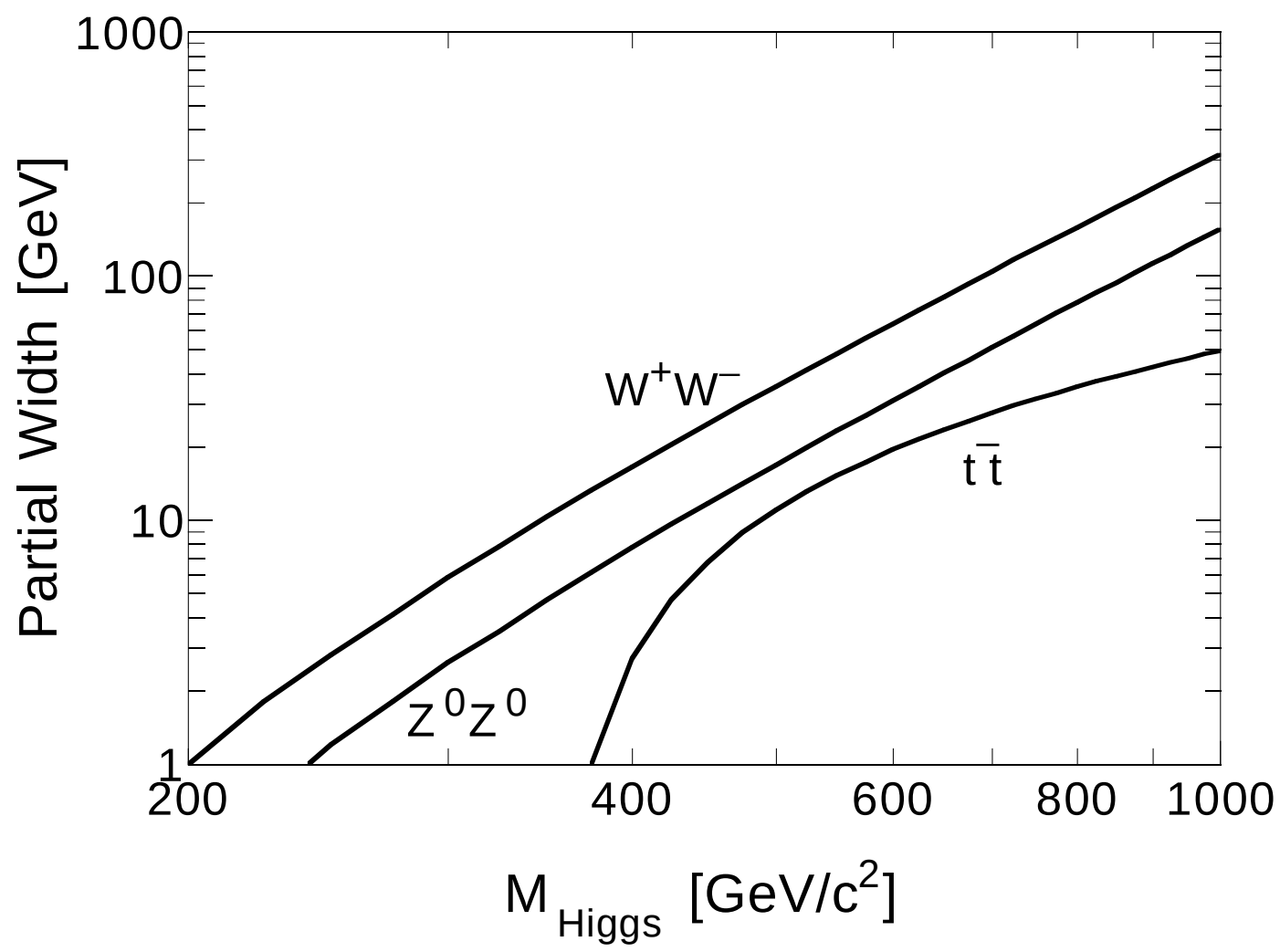

Figure 7: Partial widths for the prominent decay modes of a heavy Higgs boson.

# 9 Are We Luckier Than We Deserve?

According to the cockroach theory of stock market analysis, there is never a single piece of good news or bad news. Might the discovery of top be the precursor of other discoveries? More specifically, might the signal attributed to top also contain evidence of other new particles? Let us review two possibilities.

A heavy top influences the spectrum analysis in minimal supersymmetric extensions of the standard model. In some supersymmetric models, the lighter of the top squark eigenstates, called $\tilde{t}_1$, is less massive than the top quark: $m_{\tilde{t}_1} < m_t$ [32]. Typically, the cross section for production of a heavy squark pair in $q\bar{q}$ interactions is about 1/8 to 1/4 of the cross section for production of a pair of quarks of the same mass. Consequently, production of $\tilde{t}_1\tilde{t}_1^*$ is unlikely to distort the "top" production cross section dramatically. If it is kinematically allowed, the chain

$$\begin{array}{lll} \tilde{t}_1 & \rightarrow & b\widetilde{W}^+ \\ & & \quad \hookrightarrow W^+\tilde{\chi}^0 \\ & & \qquad \hookrightarrow \ell\nu \; , \end{array} \tag{15}$$

where $\widetilde{W}^+$ is a wino and $\tilde{\chi}^0$ a neutralino, should be prominent among many decay channels. It is a challenge to devise search strategies for the top squark and its decay products.

A second new quark, nearly degenerate with top, would have a stronger influence on "$\sigma(t\bar{t})$." Barger and Phillips [33] have explored the consequences of a weak-isoscalar, charge-2/3 quark, $t_s$, close in mass to top. This singlet quark would decay by mixing with top ($t_s \to t \to bW^+$), so would populate the same decay modes. By choosing $m_{t_s} \approx m_t$, it is easy to double the apparent top cross section. Given the close agreement between measured and calculated cross sections for top production, we do not currently have a phenomenological incentive to do this. However, the questions raised by this scenario are important and of general interest: Is the reconstructed top-quark mass distribution normal? Is the $t\bar{t}$ effective-mass distribution normal? Some numerical examples have been studied by Lane [34].

# 10 Resonances in $t\bar{t}$ Production?

Because objects associated with the breaking of electroweak symmetry tend to couple to fermion mass, the discovery of top opens a new window on electroweak symmetry breaking. The possibility of new sources of $t\bar{t}$ pairs makes it urgent to test how closely top production conforms to standard (QCD) expectations.

Two classes of models have received considerable attention in the context of a heavy top quark. Top-condensate models and multiscale technicolor both imply the existence of color-octet resonances that decay into $t\bar{t}$, for which the natural mass scale is a few hundred GeV/$c^2$. We are led to ask: Is there a resonance in $t\bar{t}$ production? How is it made? How (else) does it decay?

In the standard Higgs model of electroweak symmetry breaking, the Higgs potential breaks the $SU(2)_L \otimes U(1)_Y$ gauge symmetry. The nonzero vacuum expectation value of the Higgs field endows the $W^\pm$ and $Z^0$ with mass and, through the arbitrary Yukawa couplings of the Higgs field to fermions, gives masses to the quarks and leptons. The Higgs mechanism is a relativistic generalization of the Ginzburg–Landau phenomenology of the superconducting phase transition. Both technicolor and topcolor are dynamical symmetry breaking schemes that are inspired by the Bardeen–Cooper–Schrieffer theory of superconductivity.

In technicolor, the QCD-like technicolor interaction becomes strong at low energies and forms a technifermion condensate that breaks chiral symmetry and gives masses to the gauge bosons $W^\pm$ and $Z^0$. In a generalization of the basic scheme known as extended technicolor, new gauge bosons couple ordinary fermions to technifermions and allow the fermions to communicate with the technifermion condensate and acquire mass [35]. In topcolor, a new interaction drives the formation of a $\langle \bar{t}t \rangle$ condensate that hides the electroweak symmetry and gives masses to the ordinary fermions [36].

In the technicolor picture, which has been elaborated recently by Eichten and Lane [37], a color-octet analogue of the $\eta'$ meson, called $\eta_T$, is produced in gluon-gluon interactions. The sequence

$$gg \to \eta_T \to (gg, t\bar{t}) \qquad (16)$$

leads to distortions of the $t\bar{t}$ invariant-mass distribution, and of the two-jet invariant-mass distribution. Since $\Gamma(\eta_T \to b\bar{b})/\Gamma(\eta_T \to t\bar{t}) = m_b^2/m_t^2$, there should be only a negligible perturbation on the $b\bar{b}$ invariant-mass distribution.

In the topcolor picture explored by Hill and Parke [38], a massive vector "coloron" can be produced in $q\bar{q}$ interactions. The coloron decays at comparable rates into $t\bar{t}$ and $b\bar{b}$ and can appear as a resonance peak in both channels. There is no clear reason to expect the coloron to distort the untagged two-jet invariant-mass spectrum.

If an enhancement were to be seen in the $t\bar{t}$ channel, a useful differential diagnostic, in addition to the $b\bar{b}$ and jet-jet invariant-mass distributions, will be the $t\bar{t}$ mass spectrum in different rapidity intervals and at different energies. It is useful to recall from our discussion of top production rates in §3 that the relative $gg$ and $q\bar{q}$ luminosities at high masses change significantly between $\bar{p}p$ energies of 1.8 and 2.0 TeV. At the LHC, the large $gg$ luminosity would greatly enhance the contribution of $\eta_T$ with respect to the standard QCD process. The isotropic decays of $\eta_T$ will help to characterize the technicolor case. Hill and Parke [38] and Lane [34] have investigated the discriminatory power of $d\sigma/dp_\perp$ and of the production angular distributions for top.

When searching for resonances and other exotic sources of top, it is important not to apply cuts that efficiently exclude all mechanisms but standard QCD production. Traditional expectations for sphericity, aplanarity, and similar event-shape parameters may not be realized for new sources.

## 11 The Top Quark and the Quotidian

It is popular to say that top quarks were created in great numbers in the early moments after the big bang some fifteen billion years ago, disintegrated in a fraction of a second, and vanished from the scene until my colleagues learned to create them in the Tevatron at Fermilab. That would be reason enough to be interested in top: to learn how it helped sow the seeds for the primordial universe that has evolved into the world of diversity and change we live in. But it is not the whole story; it invests the top quark with a remoteness that hides its real importance—and overlooks the immediacy of particle physics. The real wonder is that here and now, every minute of every day, top affects the world around us. I would like to close by giving one striking example of top's influence on the everyday [39].

Consider a unified theory of the strong, weak, and electromagnetic interactions—say three-generation $SU(5)$—in which all coupling constants take on a common value, $\alpha_U$, at some high energy, $M_U$. If we adopt the point of view that the value of the coupling constant is fixed at the unification scale, then the value of the QCD scale parameter $\Lambda_{\mathrm{QCD}}$ depends on the mass of the top quark. If we evolve the $SU(3)_c$ coupling, $\alpha_s$, down from the unification scale in the spirit of Georgi, Quinn, and Weinberg [40], then the leading-logarithmic behavior is given by

$$1/\alpha_s(Q) = 1/\alpha_U + \frac{21}{6\pi}\ln(Q/M_U) \ , \qquad (17)$$

for $M_U > Q > m_t$. In the interval between $m_t$ and $m_b$, the slope $(33-2n_f)/6\pi$ (where $n_f$ is the number of active quark flavors) steepens to $23/6\pi$, and then increases by another $2/6\pi$ at every quark threshold. At the boundary $Q = Q_n$ between effective field theories with $n-1$ and $n$ active flavors, the coupling constants $\alpha_s^{(n-1)}(Q_n)$ and $\alpha_s^{(n)}(Q_n)$ must match. This behavior is shown by the solid line in Figure 8.

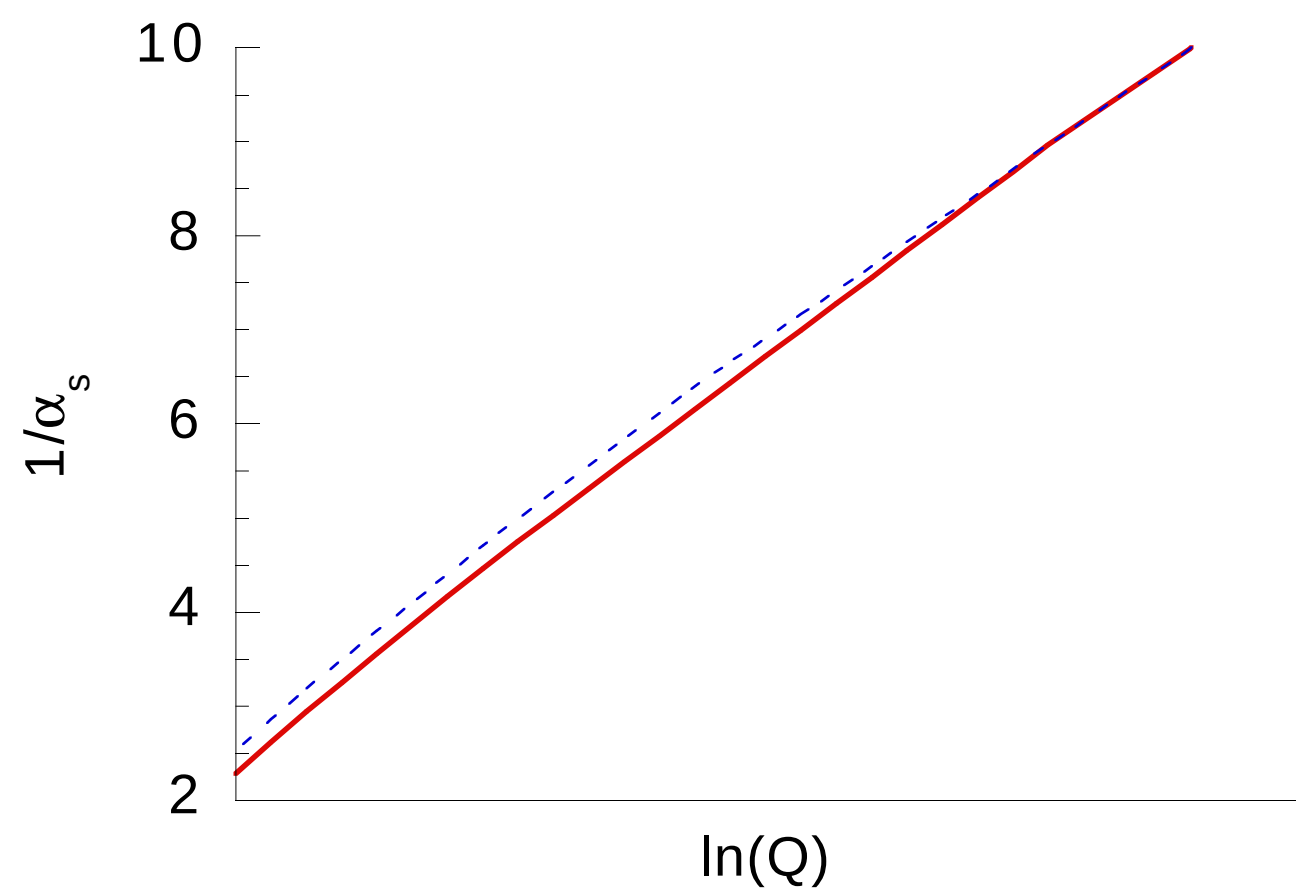

Figure 8: Two evolutions of the strong coupling constant.

To discover the dependence of $\Lambda_{\rm QCD}$ upon the top-quark mass, we use the one-loop evolution equation to calculate $\alpha_s(m_t)$ starting from low energies and from the unification scale, and match:

$$1/\alpha_U + \frac{21}{6\pi}\ln(m_t/M_U) = 1/\alpha_s(m_c) - \frac{25}{6\pi}\ln(m_c/m_b) - \frac{23}{6\pi}\ln(m_b/m_t) \quad . \tag{18}$$

Identifying

$$1/\alpha_s(m_c) \equiv \frac{27}{6\pi}\ln(m_c/\Lambda_{\rm QCD}) \quad , \tag{19}$$

we find that

$$\Lambda_{\rm QCD} = e^{-6\pi/27\alpha_U}\left(\frac{M_U}{1\ {\rm GeV}}\right)^{21/27}\left(\frac{m_t m_b m_c}{1\ {\rm GeV}}\right)^{2/27}\ {\rm GeV} \quad . \tag{20}$$

The scale parameter $\Lambda_{\rm QCD}$ is the only dimensionful parameter in QCD; it determines the scale of the confinement energy that is the dominant contribution to the proton mass. We conclude that, in a simple unified theory,

$$M_{\rm proton} \propto m_t^{2/27} \quad . \tag{21}$$

The dotted line in Figure 8 shows how the evolution of $1/\alpha_s$ changes if the top-quark mass is reduced. We see from Equations (20) and (21) that a factor-of-ten decrease in the top-quark mass would result in a 20% decrease in the proton mass. The microworld does determine the behavior of the quotidian! Let us savor the realization that we can understand the origin of one of the most important parameters in the everyday world—the mass of the proton—only by knowing the properties of the top quark.

Top matters!

## Acknowledgments

Fermilab is operated by Universities Research Association, Inc., under contract DE-AC02-76CHO3000 with the United States Department of Energy. I am grateful to the Department of Physics and Laboratory of Nuclear Studies at Cornell University for warm hospitality during the spring semester of 1995. I thank Estia Eichten, Chris Hill, Ken Lane, and Stephen Parke for many discussions of top-quark physics and for generous assistance in the preparation of this talk. I thank Bob Cahn for many stimulating discussions about the connections between particle physics and the everyday world. For helpful comments on the manuscript, I am grateful to Gustavo Burdman, Boris Kayser, Ken Lane, and Stephen Parke. Finally, it is my pleasure to salute the marvelous *carioca* hospitality of Alberto Santoro and his LAFEX colleagues.